\documentclass{article}
\usepackage{spconf,amsmath,graphicx}
\usepackage{verbatim}
\usepackage{cite}
\usepackage{amsmath}
\usepackage{amsfonts}
\usepackage{url}
\usepackage{graphicx}
\usepackage{subfigure}
\usepackage{caption}
 \usepackage{epstopdf}
\usepackage{multirow}
\usepackage{color,xcolor}
\usepackage{CJK}
\usepackage{comment}
\usepackage{booktabs}

\usepackage{amstext}

\title{Which One is Better: Assessing Objective Metrics for Point Cloud Compression}
\name{Yipeng Liu$^\ddag$, Qi Yang$^\ddag$, Yiling Xu$^\ddag$, and Zhan Ma$^\dag$}
\address{$^\ddag$Shanghai Jiaotong University,  $^\dag$Nanjing University}

\begin{document}
%
\maketitle
\begin{abstract}
\par Point cloud compression (PCC) has made remarkable achievement in recent years. In the mean time, point cloud quality assessment (PCQA) also realize gratifying development. Some recently emerged metrics present robust performance on public point cloud assessment databases. However, these metrics have not been evaluated specifically for PCC to verify whether they exhibit consistent performance with the subjective perception. In this paper, we establish a new dataset for compression evaluation first, which contains 175 compressed point clouds in total, deriving from 7 compression algorithms with 5 compression levels. Then leveraging the proposed dataset, we evaluate the performance of the existing PCQA metrics in terms of different compression types. The results demonstrate some deficiencies of existing metrics in compression evaluation.

\end{abstract}
\section{Introduction}
\label{sec:introduction}

Recently, point cloud has emerged as a promising representation format in prevalent 3D applications (e.g., autonomous driving, augmented reality), for which the point cloud compression (PCC) is of great interest for efficient service enabling in practices. Theoretically, compressing the point cloud is to determine the optimal trade-off between the bit rate and reconstruction quality, putting forward the urgent desire of precise and robust distortion measurement. The most straightforward means for point cloud distortion evaluation is directly re-using the pixel-wise MSE (mean squared error) or equivalent PSNR (peak signal-to-noise ratio) that are prevailed for measuring the image/video quality. However, point cloud is a collection of non-uniformly scattered points, potentially having superimposed impairments from both the geometry and attribute (e.g., color) components. This is largely distant from existing distortion measures of 2D image/video where only attribute (e.g., pixel intensity) variations are considered because of well-structured pixel correspondences in paired frames.

Currently, the Moving Picture Experts Group (MPEG) has applied the separable measurements of geometry and attribute in the course of PCC standardization period. For geometry part, it mainly calculates  the point-to-point (p2point)~\cite{cignoni1998metro}, or point-to-plane (p2plane)~\cite{Tian2017Evaluation} distances; while for attribute part, the PSNRyuv is used to quantify the point cloud quality. Besides the metrics which have already been applied in MPEG PCC standardization, some other metrics present better performance and robustness than p2p and PSNRyuv on public PCQA databases. For example, PCQM~\cite{meynet2020pcmd} and GraphSIM~\cite{yang2020graphsim} improve the point-wise metrics and consider both the geometry and photometric attributes of point cloud. However, the performance of these proposed metrics have not been evaluated specifically for PCC at present.

\par The reason is the lack of dataset for evaluating compression performance. Because of the difficulty in producing point cloud dataset and the manpower cost for annotating distorted point clouds, there are only a few datasets for PCQA that have been proposed, few of which can be used for compression evaluation. PointXR~\cite{Alexiou2020PointXR} and SJTU-PCQA~\cite{Yang2020TMM3DTO2D} don't involve G-PCC, V-PCC and other major compression distortions. IRPC~\cite{Javaheri2019IRPC} and LS-PCQA~\cite{liu2020LSPCQA} only contain one or several officially recommended application cases, which, however, don't consider the geometry distortion, color distortion and both of them in compression. Thus, in the newly established dataset, the compression distortions are divided into lossless-geometry (G)-lossy-attribute (A) distortion, lossy-G-lossless-A distortion, and lossy-G-lossy-A distortion.

\par Many researches have demonstrated that the existing objective metrics may predict the inconsistent visual quality with subjective perception, such as in~\cite{liu2020LSPCQA}. The point clouds with different bitrates may have different PSNRs but present the same subjective scores. In such case, the further compression is accessible. Thus, it is necessary to conduct the evaluation of the existing PCQA performance specifically for PCC.

\par The contributions of this paper are summarized as follows:
\begin{itemize}
\item We establish a dataset specifically for PCC evaluation for the first time at present. The proposed dataset considers 7 compression distortions with 5 compression levels, containing 175 distorted point clouds in total.
\item Leveraging the proposed dataset, we evaluate the performance of the existing quality assessment metrics for several PCC algorithms. The experiment demonstrates some deficiencies of existing quality assessment metrics in PCC. Aiming at the found deficiencies, we propose some potential ways to improve the existing metrics.
\end{itemize}

\par The rest of this paper is organized as follows: Section 2 reviews the current PCC algorithms and the existing PCQA metrics. Then we introduce the proposed dateset for compression evaluation in Section 3. The subjective and objective performance is evaluated and analysed in Section 4. The conclusion is summarized in Section 5.

\section{Related Work}
\label{sec:relatedwork}

\subsection{Poind Cloud Compression}

\par MPEG G-PCC~\cite{Schwarz2019GPCC} leverages 3D models for PCC, which take advantage of Octree~\cite{Schnabel2006Octree} decomposition or triangle soup to perform the geometry coding. MPEG V-PCC~\cite{Schwarz2019GPCC} conducts 3D-to-2D projection first, and then the successful video codecs, e.g., High-Efficiency Video Coding (HEVC)~\cite{Sullivan2013HEVC}, can be used to encode the attributes, depth maps and occupation maps. Since G-PCC and V-PCC are the popular compression standardization, we evaluate the performance of the existing quality assessment metrics for G-PCC and V-PCC compression distortions in geometry and color attributes.

\subsection{Poind Cloud Quality Assessment}

\par The point-wise metrics are first applied for point cloud quality assessment. The p2point~\cite{cignoni1998metro} and p2plane~\cite{Mekuria2016Evaluation} measure the geometry error point-wisely to quantify the quality score of point cloud. The color distortion can also be measured point-wisely to quantify the distortion for geometric matched point pairs. All of these are adopted into the MPEG PCC technology standardization for compression efficiency measurement~\cite{MPEGSoft}.

\par Alexiou et al.~\cite{alexiou2018pointt} and Yang et al.~\cite{Yang2020TMM3DTO2D} propose to project 3D point cloud to 2D planes, by which classical image objective quality metrics can be applied, but such metrics depend heavily on the performance of projection method.

\par Meynet et al.~\cite{meynet2019pcmsdm} leverage curvature and lightness to improve the accuracy of quality prediction, while Viola et al.~\cite{viol2020acolor} use color histogram to quantify the distortion. Alexiou et al.~\cite{alexiou2018pointt} propose to leverage SSIM~\cite{wang2004image} to regress four attributes of point cloud into quality scores. Yang et al.~\cite{yang2020graphsim} propose the graph representation of point cloud to conduct the distortion measurement. Liu et al.~\cite{liu2020LSPCQA} propose to leverage sparse CNN to predict the quality scores of point cloud.

\par The existing quality assessment metrics are not evaluated specifically for PCC. Thus, in this work, we conduct the evaluation to verify whether the performance of existing objective metrics are consistent with the subjective perception in PCC.

\section{Dataset Construction}
\label{sec:dataset}

\par
\begin{figure}[htbp]
	\centering
	\includegraphics[width=1\linewidth]{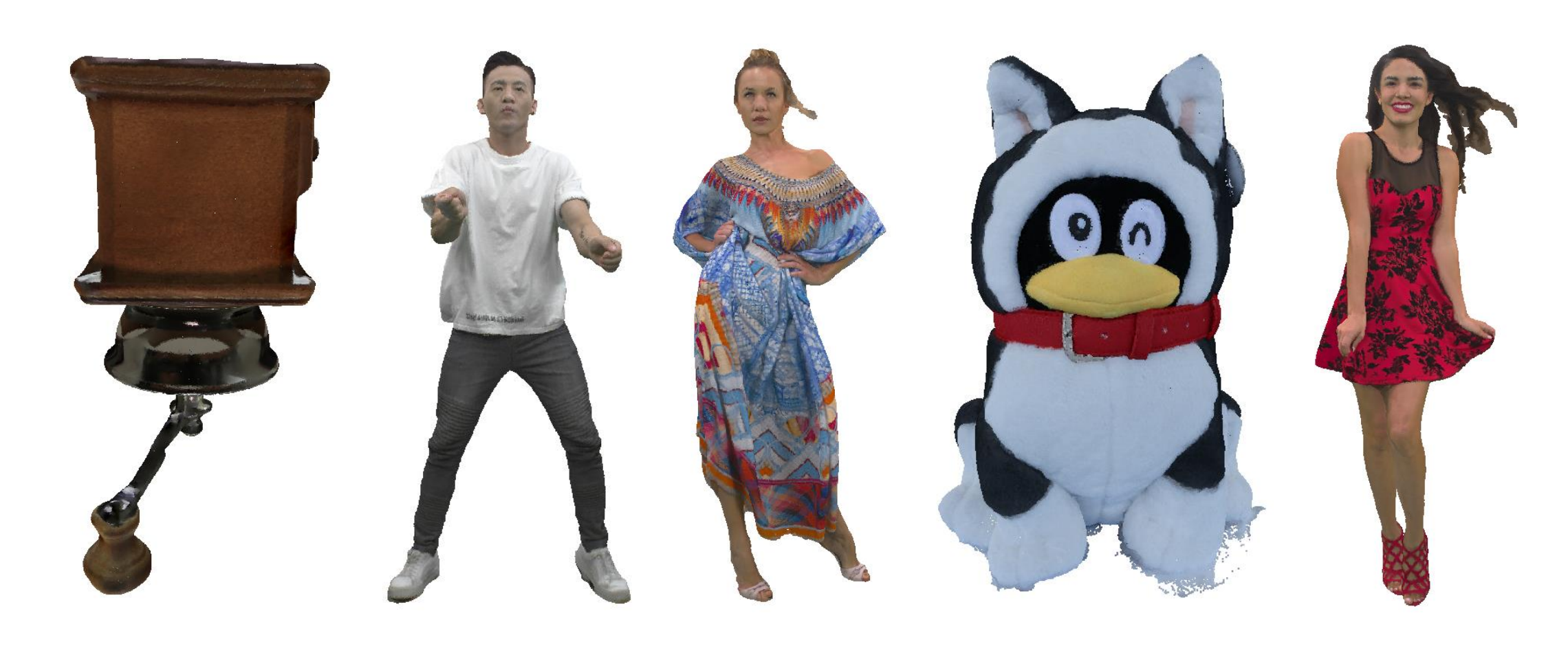}
	\caption{Snapshots of the reference point clouds.}
	\label{reference}

\end{figure}

 To better understand the compression distortion, we first establish a dataset for compression evaluation. The reference point clouds are selected from MPEG and AVS point cloud datasets, whose snapshots are shown in Fig.\ref{reference}. The reference point clouds are then distorted with 10 compression algorithms, which are shown in Table \ref{alldistortions}. We control the quantization parameters for G-PCC and V-PCC and the resolution for Octree compression to generate different levels of compression distortions.

\begin{table}[t]
\caption{Compression algorithms used for distorted point cloud generation.}
\resizebox{\columnwidth}{!}
  {
\begin{tabular}{lllll}
\hline
                                                                              & Distortions                                                                                                                 & Levels                                                                                                                                                                &  &  \\ \hline
\begin{tabular}[c]{@{}l@{}}Geometry\\ Distortion\end{tabular}                 & \begin{tabular}[c]{@{}l@{}}GPCC-lossy-geom-lossless-attrs\\      \\ VPCC-lossy-geom-lossless-attrs\\ \\ Octree\end{tabular} & \begin{tabular}[c]{@{}l@{}}(positionQuantizationScale)\\ 0.75 0.5 0.25 0.125 0.0625\\ (geomQP) \\ 22 32 37 42 51\\ (resolution)\\ 8 10 12 14 16\end{tabular}                &  &  \\ \hline
\begin{tabular}[c]{@{}l@{}}Color\\ Distortion\end{tabular}                    & \begin{tabular}[c]{@{}l@{}}GPCC-lossless-geom-lossy-attrs\\      \\ VPCC-lossless-geom-lossy-attrs\end{tabular}             & \begin{tabular}[c]{@{}l@{}}(qp) \\ 35 39 43 47 51\\ (textureQP) \\ 32 37 42 47 51\end{tabular}                                                                              &  &  \\ \hline
\begin{tabular}[c]{@{}l@{}}Both Geometry \\ and Color Distortion\end{tabular} & \begin{tabular}[c]{@{}l@{}}GPCC-lossy-geom-lossy-attrs\\      \\ VPCC-lossy-geom-lossy-attrs\end{tabular}                   & \begin{tabular}[c]{@{}l@{}}(positionQuantizationScale,qp)   \\ 0.75,35 0.5,39 0.25,43 0.125,47 0.0625,51\\ (geomQP,textureQP) \\ 24,32 28,37 32,42 36,47 40,51\end{tabular} &  &  \\ \hline
\end{tabular}
}
\label{alldistortions}%
\end{table}

\begin{table}[htbp]
  \centering
  \caption{SROCC of the objective metrics in terms of different compression types. The results highlighted in red demonstrate the additional geometry distortion introduced in V-PCC.}
  \resizebox{\columnwidth}{!}
  {
    \begin{tabular}{p{8.78em}|c|c|c|c|c|c|c|c|}
    \toprule
    \multicolumn{1}{c|}{\multirow{2}[4]{*}{}} & \multicolumn{1}{p{5em}|}{mseF,PSNR\newline{} (p2point)   } & \multicolumn{1}{p{4.89em}|}{mseF,PSNR\newline{} (p2plane)} & \multicolumn{1}{p{4.835em}|}{h.,PSNR  \newline{} (p2point)} & \multicolumn{1}{p{4.5em}|}{h.,PSNR   \newline{}(p2plane)} & PSNRyuv & h.PSNRyuv & PCQM  & \multicolumn{1}{c}{GraphSIM} \\
\cmidrule{2-9}    \multicolumn{1}{c|}{} & SROCC & SROCC & SROCC & SROCC & SROCC & SROCC & SROCC & \multicolumn{1}{c}{SROCC} \\
    \midrule
    VPCC-lossless-\newline{}geom-lossy-attrs & \textcolor[rgb]{ 1,  0,  0}{-0.485311 } & \textcolor[rgb]{ 1,  0,  0}{-0.426913 } & \textcolor[rgb]{ 1,  0,  0}{-0.239635 } & \textcolor[rgb]{ 1,  0,  0}{-0.239635 } & 0.278818  & -0.397297  & 0.186011  & \multicolumn{1}{c}{0.152047 } \\
    \midrule
    VPCC-lossy-\newline{}geom-lossless-attrs & 0.310745  & 0.317579  & 0.294263  & 0.523805  & 0.447425  & 0.099696  & 0.356975  & \multicolumn{1}{c}{0.132660 } \\
    \midrule
    VPCC-lossy-\newline{}geom-lossy-attrs & 0.214142  & 0.218480  & 0.367945  & 0.502424  & 0.419607  & -0.432621  & 0.406199  & \multicolumn{1}{c}{0.420790 } \\
    \midrule
    GPCC-lossless-\newline{}geom-lossy-attrs & -     & -     & -     & -     & 0.820883  & 0.927501  & 0.820883  & \multicolumn{1}{c}{0.828098 } \\
    \midrule
    GPCC-lossy-\newline{}geom-lossless-attrs & 0.888581  & 0.899010  & 0.912863  & 0.888853  & 0.819330  & 0.834348  & 0.234035  & \multicolumn{1}{c}{0.881489 } \\
    \midrule
    GPCC-lossy-\newline{}geom-lossy-attrs & 0.923528  & 0.914763  & 0.932438  & 0.922650  & 0.866554  & 0.736273  & 0.949425  & \multicolumn{1}{c}{0.932692 } \\
    \midrule
    Octree & 0.856948  & 0.805531  & 0.880016  & 0.835823  & 0.686355  & -0.059388  & 0.819083  & 0.810713  \\
    \bottomrule
    \end{tabular}%
    }
  \label{SROCC}%
\end{table}%

\par To annotate the proposed dataset, a subjective experiment is organized to collect Mean Opinion Scores (MOSs). Since double stimulus method can obtain more stable results for minor distortion, we adopt double stimulus method for annotating the distorted point clouds in this work. The reference and distorted point clouds are presented successively. Then the candidate gives the score of 1-5 based on the impairment scale, where score 5 denotes imperceptible impairment and score 1 denotes very annoying impairment. We convene 32 candidates to score the 175 distorted point clouds to ensure at least 16 subjective scores for each distorted point cloud after outlier removal. The experiment steps and environment strictly follow ITU-R Recommendation BT. 500~\cite{BT500}. Such method for collecting subjective MOSs is also adopted in~\cite{Wu20206Dof,Su2019PQA}.

\section{Evaluation}
\label{sec:experiment}

\subsection{Subjective comparison under different QPs}

\par
\begin{figure}[htbp]
	\centering
	\includegraphics[width=0.75\columnwidth]{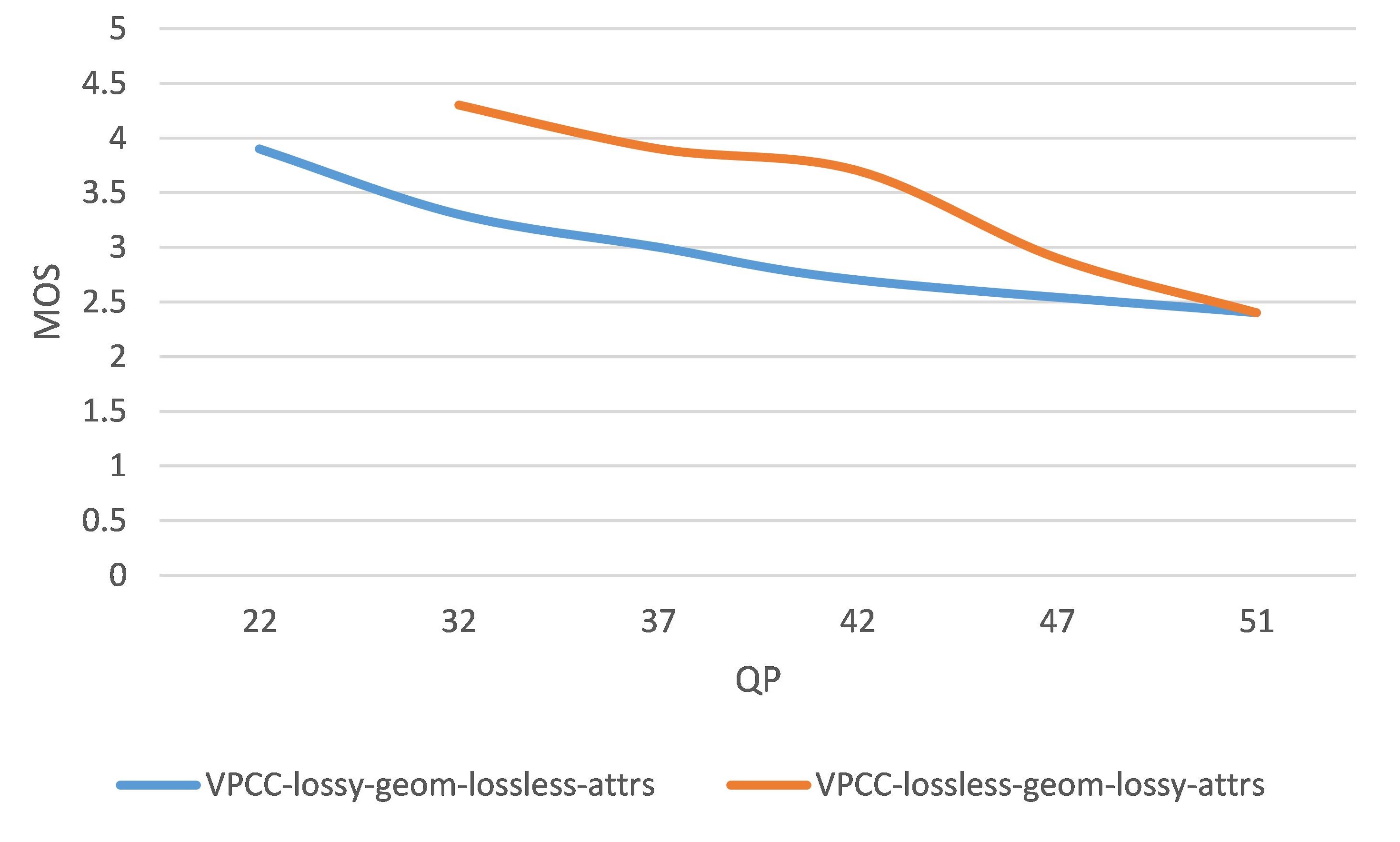}
	\caption{MOS comparison for V-PCC geometry and color distortion at different QPs.}
	\label{MOScomparison}

\end{figure}

 Through the subjective experiment, we study the sensitivity of human eye to geometry and attribute compression. In Fig.\ref{MOScomparison}, we compare the MOSs of distorted point clouds under different quantization steps (QPs) for V-PCC geometry and color distortion. The subjective experiment demonstrates that the participants can easily recognize the geometry distortions. It can be seen from Fig.\ref{MOScomparison} that when geometry compression exists, the point clouds exhibit relative lower MOSs. But for compression of different levels, the human visual system (HVS) is more sensitive to the QP changes in color compression. It demonstrates that to predict the quality scores, metrics should quantify the color impairments based on considering the geometry distortion.

\subsection{Objective evaluation for different compressions}

\par Leveraging the proposed dataset, we evaluate the listing compression distortions using several objective metrics with details in Table~\ref{detail}.

\begin{table}[htbp]
\scriptsize
  \centering
  \caption{Comparison of the metrics used for evaluation.}
    \begin{tabular}{l|l}
    \toprule
    p2point~\cite{cignoni1998metro} & Symmetric rms distance between the point clouds  \\
    \midrule
    p2plane~\cite{Mekuria2016Evaluation} & \multicolumn{1}{p{22.22em}}{Symmetric rms distance between the clouds with \newline{}the normal vector as weights} \\
    \midrule
    Hausdroff p2point~\cite{cignoni1998metro} & Symmetric Haussdorf distance between the clouds  \\
    \midrule
    Hausdroff p2plane~\cite{Mekuria2016Evaluation} & \multicolumn{1}{p{22.22em}}{Symmetric Haussdorf distance between the clouds \newline{}with the normal vector as weights} \\
    \midrule
    PSNRyuv~\cite{MPEGSoft} & \multicolumn{1}{p{22.22em}}{Peak signal to noise ratio derived from rms distance \newline{}for YUV between the clouds } \\
    \midrule
    Hausdroff PSNRyuv~\cite{MPEGSoft} & \multicolumn{1}{p{22.22em}}{Peak signal to noise ratio derived from Haussdorf \newline{}distance for YUV between the clouds } \\
    \midrule
    PCQM~\cite{meynet2020pcmd}  & \multicolumn{1}{p{22.22em}}{Optimally-weighted linear combination of \newline{}curvature and color lightness features} \\
    \midrule
    GraphSIM~\cite{yang2020graphsim} & SSIM form of point cloud color gradient  \\
    \bottomrule
    \end{tabular}%
  \label{detail}%
\end{table}%

\par We use Spearman Rank Order Correlation Coefficient (SROCC) to measure the correlation between the objective quality scores and subjective MOSs. A quality assessment metric is considered to have the best performance when SROCC is close to 1. The SROCCs of these metrics for each compression distortion based on the subjective MOSs are listed in Table \ref{SROCC}.

\par The p2point, p2plane and PSNR measure the geometry or color impairment to predict the quality scores, which violates the fusion evaluation of HVS. We can see from last subsection that the geometry and color attributes should both play important roles in HVS.

\par
\begin{figure}[htbp]
	\centering
	\includegraphics[width=1\linewidth]{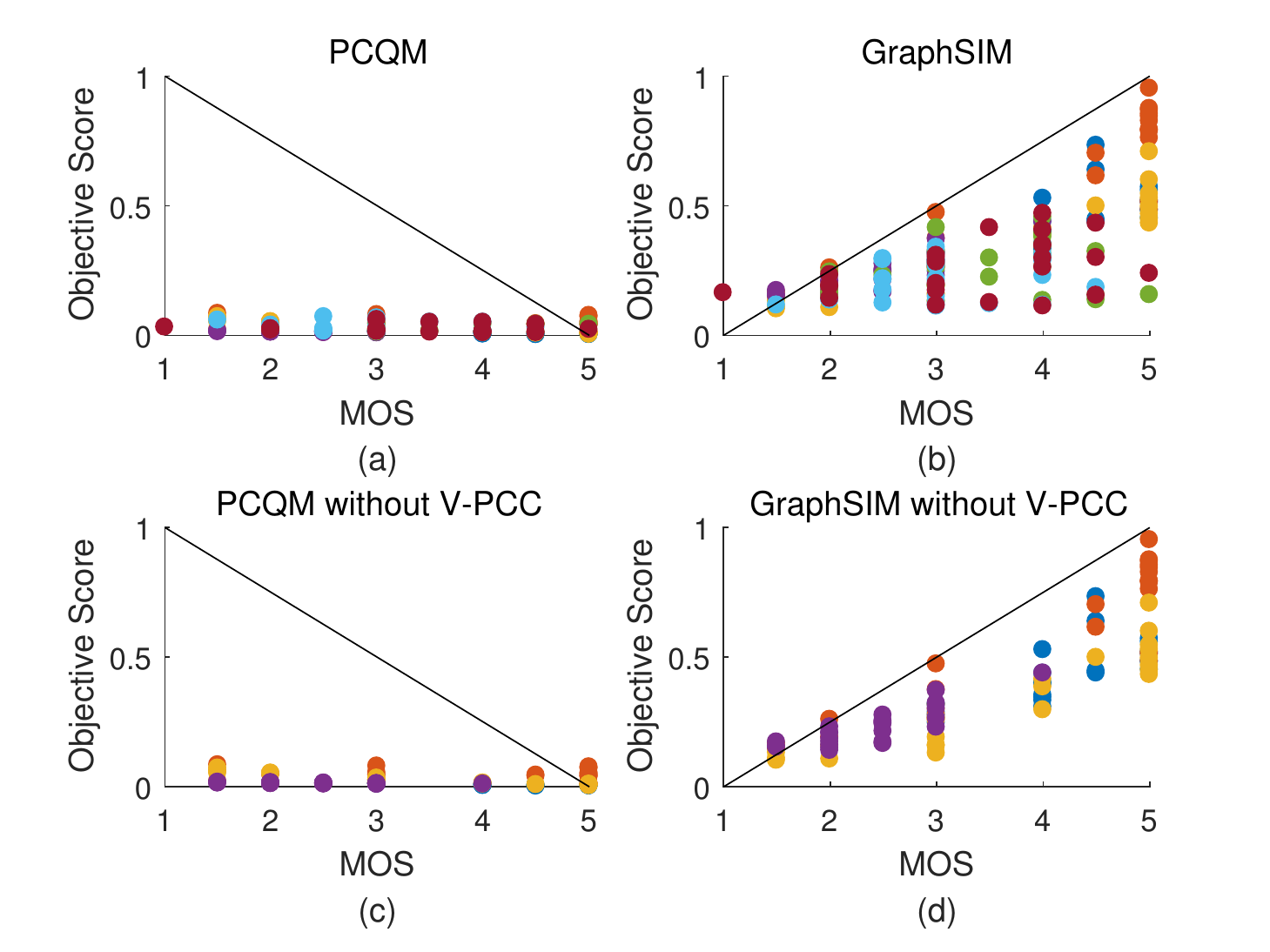}
	\caption{MOS prediction accuracy of objective metrics: subplots (a)(b) are results of PCQM and GraphSIM for compression distortions; subplots (c)(d) remove the results for V-PCC distortions; solid lines imply the perfect prediction for PCQM and GraphSIM respectively when being overlapped with this line. }
	\label{plot}

\end{figure}

 To elaborate consistency between the metrics for superimposed impairments and HVS, we provide the scatter plots shown in Fig.\ref{plot} for PCQM and GraphSIM. Although PCQM exhibit consistent monotonicity with subjective MOSs which can be seen in Table \ref{SROCC}, Fig.\ref{plot} demonstrates that PCQM is insensitive to the impairments in compression distortions. For all the distorted samples, PCQM predicts the relatively high quality. GraphSIM exhibits more consistent performance with HVS except for V-PCC distortions. We list some samples of V-PCC in Fig.\ref{examples} and discuss the reasons for the poor performance in V-PCC in Section 4.3. Besides, the results demonstrates that GraphSIM predicts different objective scores for point clouds with similar MOSs, which comes from the inconsistency with HVS. The existing metrics predict the quality scores by measuring the impairment of geometry attributes or color attributes or both of them. However, in many cases, the inflexible assessment metrics violate the HVS scoring.

\subsection{Performance comparison of whether samples are from MPEG 8i dataset}
\label{sec:discussion}

\par We can see from Table \ref{SROCC} that all the listing metrics exhibit poor performance for V-PCC distortions. To find out the causes for the unsatisfactory objective performance in V-PCC, we compute the SROCC for samples from MPEG 8i dataset and samples from other datasets respectively. The results are shown in Table \ref{SROCC2}. We notice that the existing quality assessment metrics have normal performance for samples from MPEG 8i dataset but exhibit poor performance for other point clouds.

\par
\begin{figure}[htbp]
	\centering
	\includegraphics[width=1\linewidth]{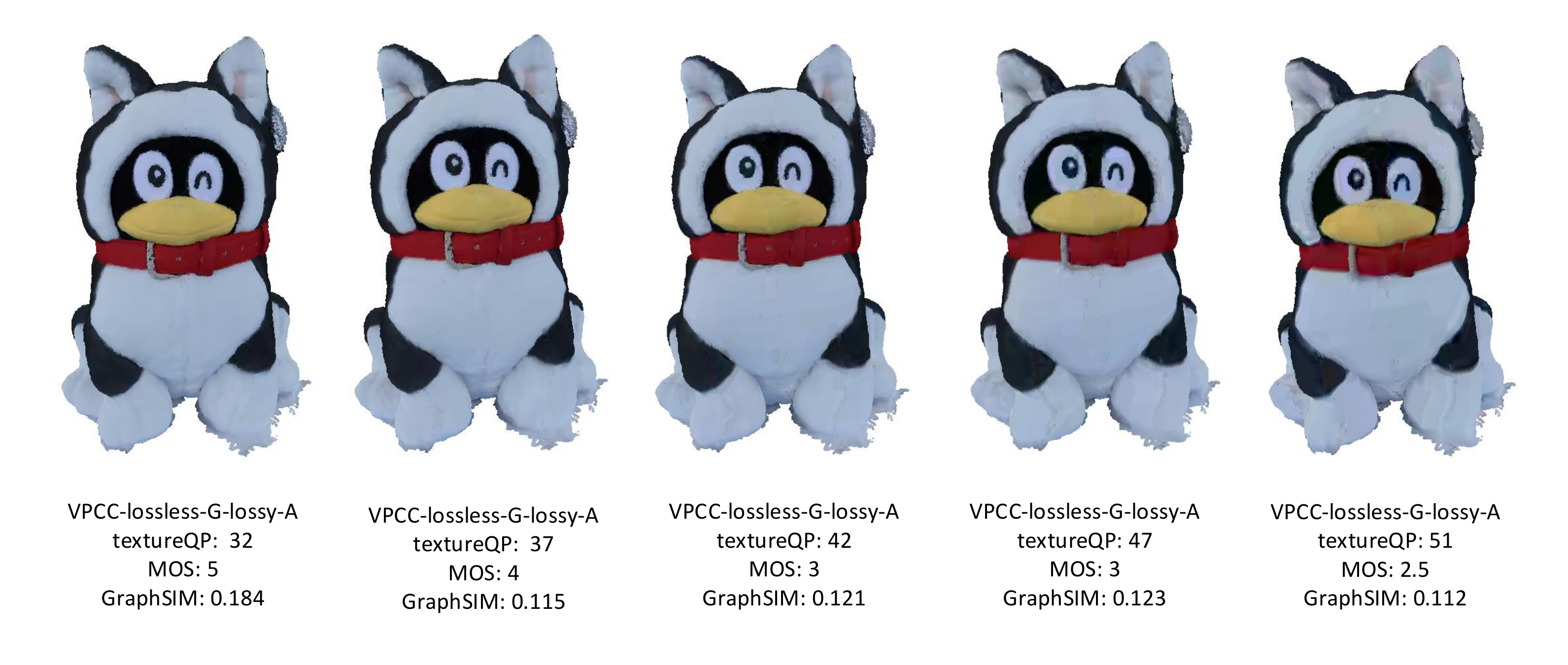}
	\caption{Some scoring examples in V-PCC. }
	\label{examples}

\end{figure}

\begin{figure}[htbp]
	\centering
	\includegraphics[width=1\linewidth]{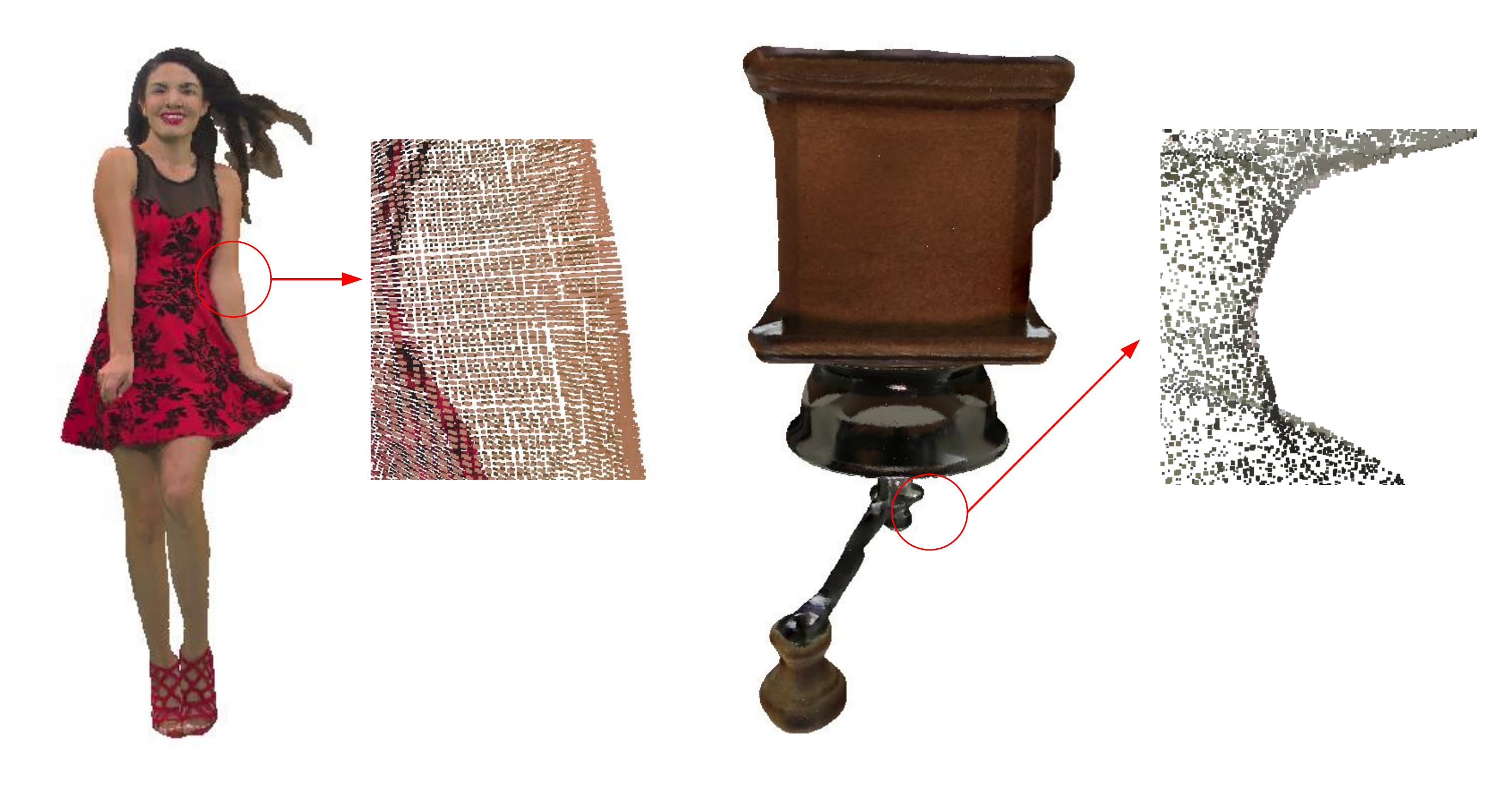}
	\caption{The spatial distribution comparison between samples from MPEG 8i dataset and other datasets. }
	\label{uniform}

\end{figure}

 The different is that the point clouds in MPEG 8i dataset is voxelized and post-processed to scattered on a regular 3D grid~\cite{Eon2017MPEG8i}, while the spatial distribution of other raw point clouds is not uniform, which can be seen in Fig.\ref{uniform}. For raw point clouds, the projection in V-PCC may induce the additional geometry distortions, which interferes with the performance of the existing quality assessment metrics. The results in Table \ref{SROCC} which are highlighted in red demonstrate
the additional geometry distortion introduced in V-PCC. For example, the Cartesian coordinate of point may change after compression, and the number of points may increase dramatically. We can see from Fig.\ref{examples} that GraphSIM gives low quality scores for samples of V-PCC distortion with high MOSs. The induced additional distortions result in the much lower objective scores compared with the subjective perception.

\subsection{Potential ways for improvement}

\par Since the structure impairment is more obvious in HVS, the geometry distortions in objective metrics deserve greater weights.

\par To address the mentioned problems, we can build the specific nonlinear regressor to map the geometric or color errors to the predicted quality scores. The metrics may respond to the errors within the set threshold to emulate the limit resolution of HVS. Besides, the key point sampling may be adopted. By improving the key point sampling algorithm, the interference on metric performance derived from compression may be mitigated.

\begin{table}[htbp]
\tiny
  \centering
  \caption{SROCC of samples in MPEG 8i dataset and samples from other datasets.}
    \begin{tabular}{p{8.39em}|c|c|p{8.39em}|c|c}
    \toprule
    \multicolumn{1}{c|}{\multirow{2}[4]{*}{Samples in 8i}} & PCQM  & GraphSIM & \multicolumn{1}{c|}{\multirow{2}[4]{*}{Samples not in 8i}} & PCQM  & GraphSIM \\
\cmidrule{2-3}\cmidrule{5-6}    \multicolumn{1}{c|}{} & SROCC & SROCC & \multicolumn{1}{c|}{} & SROCC & SROCC \\
    \midrule
    VPCC-lossless-\newline{}geom-lossy-attrs & 0.797821  & 0.592133  & VPCC-lossless-\newline{}geom-lossy-attrs & -0.108059  & -0.069921  \\
    \midrule
    VPCC-lossy-\newline{}geom-lossless-attrs & 0.779246  & 0.570180  & VPCC-lossy-\newline{}geom-lossless-attrs & 0.373700  & 0.177016  \\
    \midrule
    VPCC-lossy-\newline{}geom-lossy-attrs & 0.876694  & 0.722346  & VPCC-lossy-\newline{}geom-lossy-attrs & 0.152048  & 0.196395  \\
    \bottomrule
    \end{tabular}%
  \label{SROCC2}%
\end{table}%

\section{Conclusion}
\label{sec:conclusion}

\par In this paper, we proposed a dataset for compression evaluation derived from 5 selected point clouds with 7 compression distortions at 5 compression levels. The proposed dataset has 175 distorted point clouds in total for evaluating compression. Leveraging the proposed dataset, we evaluated the performance of the existing PCQA metrics in terms of different compression types. The experiment demonstrates that the existing quality assessment metrics cannot measure the point cloud quality for PCC in consistency with human perception. Besides, based on the experiment result, we explained the shortcomings of existing metrics and provided some potential ways to improve the quality prediction for PCC.



\bibliographystyle{IEEEbib}
\bibliography{strings,refs,ref_ICIP}

\end{document}